\documentclass[%
reprint,
superscriptaddress,
amsmath,amssymb,
aps,
prapplied,
floatfix,
]{revtex4-2}

\usepackage[english]{babel}
\usepackage[normalem]{ulem}
\usepackage{amsmath} 
\usepackage{amssymb} 
\usepackage{environ} 
\usepackage{textcomp}
\usepackage{import}
\usepackage{multirow}
\usepackage{nicefrac}
\usepackage{xcolor}
\usepackage{graphicx}
\usepackage{bm}
\usepackage{upgreek}
\usepackage{booktabs}

\begin{document}

\pagestyle{plain}

\title{Energy-efficient stochastic computing  with superparamagnetic tunnel junctions}

\author{Matthew W. Daniels}
\email{matthew.daniels@nist.gov}
\affiliation{Physical Measurement Laboratory, National Institute of Standards and Technology, Gaithersburg, MD, USA}
\affiliation{Institute for Research in Electronics and Applied Physics, University of Maryland, College Park, MD, USA}

\author{Advait Madhavan}
\affiliation{Physical Measurement Laboratory, National Institute of Standards and Technology, Gaithersburg, MD, USA}
\affiliation{Institute for Research in Electronics and Applied Physics, University of Maryland, College Park, MD, USA}

\author{Philippe Talatchian}
\affiliation{Physical Measurement Laboratory, National Institute of Standards and Technology, Gaithersburg, MD, USA}
\affiliation{Institute for Research in Electronics and Applied Physics, University of Maryland, College Park, MD, USA}

\author{Alice Mizrahi}
\affiliation{Physical Measurement Laboratory, National Institute of Standards and Technology, Gaithersburg, MD, USA}
\affiliation{Institute for Research in Electronics and Applied Physics, University of Maryland, College Park, MD, USA}
\affiliation{Unit\'e Mixte de Physique, CNRS, Thales, Univ. Paris-Sud, Universit\'e Paris-Saclay, 91767 Palaiseau, France}

\author{Mark D. Stiles}
\email{mark.stiles@nist.gov}
\affiliation{Physical Measurement Laboratory, National Institute of Standards and Technology, Gaithersburg, MD, USA}

\begin{abstract}

Superparamagnetic tunnel junctions (SMTJs) have emerged as a competitive, realistic nanotechnology to support novel forms of stochastic computation in CMOS-compatible platforms. One of their applications is to generate random bitstreams suitable for use in stochastic computing implementations. We describe a method for digitally programmable bitstream generation based on pre-charge sense amplifiers.  {This generator is significantly more energy efficient than SMTJ-based bitstream generators that tune probabilities with spin currents and a factor of two more efficient than related CMOS-based implementations. The true randomness of this bitstream generator allows us to use them as the fundamental units of a novel neural network architecture.} To take advantage of the potential savings, we codesign the algorithm with the circuit, rather than directly transcribing a classical neural network into hardware. The flexibility of the neural network mathematics allows us to adapt the network to the explicitly energy efficient choices we make at the device level. The result is a convolutional neural network design operating at $\approx 150$~nJ per inference with 97~\% performance on MNIST---{a factor of $1.4$ to $7.7$ improvement in energy efficiency over comparable proposals in the recent literature.}
\end{abstract}

\maketitle

\section{Introduction}

Magnetic tunnel junctions are poised to make significant contributions to new computer chips, most immediately from non-volatile memory applications \cite{apalkov2016magnetoresistive,hanyu2016standby, guo2013ac}, such as magnetic random access memory (MRAM).  These devices consist of two magnetic layers separated by a thin tunnelling barrier.  The memory values of \texttt{0} and \texttt{1} are encoded in two different stable configurations of the device (parallel and anti-parallel magnetizations).  Values can be read by passing a small current ($\approx$~10~{\textmu}A) through the device, since its resistance depends on its configuration. The device state can be switched by overcoming an energy barrier between the two configurations by passing a higher current through the device. Since MRAM is used as a non-volatile memory, retention requirements demand that the energy barrier be kept high (greater than 40 $kT$).


However, if the energy barrier is decreased by a factor of ten (to $\approx 4\,kT$) \cite{bapna2016magnetostatic, mizrahi2018neural, rippard2011thermal}, then thermal fluctuations at room temperature cause the device to randomly switch between its stable configurations. In this case, the mean time between thermal switching events is about 55~ns, and the magnetic tunnel junction is said to be in a superparamagnetic state, making such devices superparamagnetic tunnel junctions (SMTJs). In principle, the barrier could be lowered even further for faster switching.  The relative times spent in each configuration can be controlled by passing a current through the device creating a spin-transfer torque \cite{sun2008magnetoresistance} or by passing a current through an adjacent heavy metal layer creating a spin-orbit torque \cite{lee2016emerging}. They can currently be fabricated down to a $10\;\text{nm}$ length scale~\cite{sato2017magnetic,perrissin2019perpendicular}.

The low energy, truly random behavior, ease of control, and established compatibility with complementary-metal-oxide-semiconductor (CMOS) circuitry has led to the use of SMTJs as the basis for a number of novel computing schemes \cite{grollier2016spintronic,roy2018perspective,camsari2019p}. SMTJs were proposed to implement the concept of probabilistic bits, or $p$-bits, which were leveraged for applications as Bayesian neural networks \cite{faria2018implementing,jia2018spintronics,sengupta2016probabilistic}, invertible Boolean logic \cite{PhysRevX.7.031014,borders2019integer}, reservoir computing \cite{ganguly2017reservoir}, and Ising network models applied to optimization problems \cite{hassan2019voltage, shim2017ising}. SMTJs were also proposed as stochastic neural units \cite{sengupta2016magnetic} that can interact with synaptic units, emulated by crossbar arrays of magnetic tunnel junctions that can switch stochastically in the presence of current pulses~\cite{vincent2015spin, srinivasan2016magnetic}. 

Some of these schemes encode information in the switching rates \cite{mizrahi2018neural} and others encode information in the relative time spent in each state \cite{PhysRevX.7.031014,camsari2019p}.  In these cases, the rates or probabilities are controlled by currents. Though current-controlled methods have been used in many previous device proposals, the ohmic losses they incur can be significant.

Many of these previous applications can be classified as types of probabilistic computing. The specific (and ambiguously named) subfield of probabilistic computing we consider in this paper is called \emph{stochastic computing}~\cite{gross2019stochastic}. Many of the works mentioned above would not qualify as stochastic computing \emph{per se}.  Stochastic computing is concerned with encoding real-valued numbers as the expectation values of random bitstreams; for example, a bitstream such as \texttt{0100110100} has four ones and six zeros, thereby encoding the value 0.4, since the probability of seeing a \texttt{1} on this wire is 4/10.  Stochastic computing is a competitive candidate for energy-efficient application specific architectures~\cite{alaghi2018promise}. Its robustness to noise~\cite{qian2010architecture}, high density~\cite{Lee_2018, qian2010architecture}, intrinsic parallelism~\cite{alaghi2018promise,qian2010architecture}, and latency-precision trade-off~\cite{alaghi2018promise} make it a promising platform for the implementation of dataflow-based computations in CMOS circuits. 

Ideally, stochastic computers operate on long chains of non-repeating, uncorrelated bitstreams that are cheap to produce from an area and energy efficiency standpoint. The conventional way of producing bitstreams is based on a circuit called the linear feedback shift register (LFSR), which comprises a series of flip-flops and simple combinational circuits. LFSRs operate by cycling through all of their internal binary states, each producing a pseudorandom bit, before returning back to the initial state. When LFSRs are used to generate a vast number of pseudorandom bitstreams, those bitstreams are both periodic and cross-correlated. Such correlations are usually undesirable from a computational perspective~\cite{neugebauer2018s,alaghi2018promise}. Pseudorandom bitstream generation can incur significant overheads in accelerator architectures~\cite{de2015exploring,Lee_2018}. 

Stochastic bitstreams with neither periodicity nor cross-correlation can be generated by SMTJs. The thermal nature of their switching behaviour makes bitstreams generated by SMTJ circuits aperiodic and truly random~\cite{fukushima2014spin,PhysRevApplied.8.054045,parks2018superparamagnetic}. We replace LFSR-based stochastic sources with arrays of SMTJs that use energy-efficient readout circuitry and programmable logic to generate truly random bitstreams, demonstrating their application in a convolutional neural network architecture. 

{
Using SMTJs to generate stochastic bitstreams becomes useful only when the statistics of those bitstreams can be controlled. Previous works have focused on using current biasing to realize a steady-state spin-transfer torque on the free layer of the junction. This modifies the effective energy landscape of the device so that one of the configurations is tunably preferred over the other. The use of spin-orbit torques in a similar context has also been explored. In this paper, however, we demonstrate how statistical control can be readily accomplished by traditional circuit design; moreover, we show that such digital control has superior energy efficiency over current control for a large range of reasonable material parameters. We discuss the challenges to be overcome, and the contexts to be used, wherein spintronic probability control may become an efficient option.}

{To make this demonstration concrete, we develop the circuits and architecture to use SMTJs as the basis for a neural network designed to recognize hand-written digits. This allows us to compare, in an application, digitally programmable SMTJ-sourced stochastic bitstreams against spintronically controlled versions.  Considering a full scale stochastic computing application also lets us compare SMTJ-sourced stochastic bitstreams against the performance of digitally generated pseudorandom bitstreams.} The interplay between high level design requirements for the neural network and the capabilities of the low level devices (SMTJs) leads to modifications in the design of both the networks and the circuits that connect with the devices.  Such engineering across the computational hierarchy, or \emph{stack}, plays an important role throughout this paper. In traditional computational systems, cross-stack engineering is not necessary because clean abstraction layers have been identified between different levels of design to allow optimization of each layer by itself. Computer programmers do not need to know the details of circuits in order to write code, and electrical engineers do not need to know the details of device physics in order to layout useful circuits. These clean abstraction layers break down when using novel devices or novel architectures.  

In our case, for example, implementing the neural network with stochastic bitstreams requires uncorrelated bits that can be generated at low energy, a requirement at the device and circuit level driven by the architecture.  At the same time, the use of simple primitives, \texttt{AND} and \texttt{OR} gates, to implement pieces of the neural network dictates changes to the high level structure of the neural network.  Similar engineering across the computational stack is important for the many bio-inspired or other alternative computing approaches that aim to take advantage of materials and devices wherein the native dynamics manifest the behavior of neural processes. Because the algorithmic operation depends on the physics (but not vice versa), bottom-up approaches like ours, which choose interesting or energy-efficient physical systems at the foundational level, demand that we pay attention to whether and how the high-level computation can effectively utilize the physics.

In Sec.~\ref{sec:smtj-pcsa-readout}, we propose a circuit based on a precharge sense amplifier (PCSA) to read the states of SMTJs to generate a stochastic bitstream.  Including a set-reset latch with the PCSA fixes the output to a form useful for stochastic computing. This modified PCSA avoids controlling the state of the SMTJ by write currents; these currents are much higher than the currents needed to read the state, and give rise to ohmic losses that can dominate the energy consumption of the device.  Since the expected value of the bitstream is fixed in the absence of tunable write currents, multiple such bitstreams must be combined to produce variable expected values.  In Sec.~\ref{sec:bitstream-gen}, we show how to use digital logic to combine these low energy SMTJ-based stochastic oscillators into programmable bitstream generators. These circuits operate at lower energies than LFSRs and other approaches based on SMTJs.  

In Sec.~\ref{sec:neuralnetworks}, we design and simulate a deep convolutional neural network based on LeNet5~\cite{lecun1998gradient}  to demonstrate the effectiveness of this SMTJ-based approach.  The true randomness of SMTJs relaxes constraints on design space considerations for stochastic circuits, allowing us to take advantage of uncommon stochastic computing ideas. We choose an architecture that uses logical \texttt{OR} gates as neurons, minimizing area and power expenditure compared to state-machine based approaches. The \texttt{OR} gate simultaneously provides both the summation and nonlinear activation function of a neuron.  The use of these devices saves enough energy to justify the modifications needed in the high level architecture. We train the stochastic neural network by backpropagation on an analytic approximation to the network. {In Sec.~\ref{sec:eval}, we give the results for the accuracy and energy efficiency of this approach based on simulations of this network architecture.} 

\section{Precharge sense amplifier readout of superparamagnetic tunnel junctions}
\label{sec:smtj-pcsa-readout}

\begin{figure}
    \centering
    \includegraphics[width=\columnwidth]{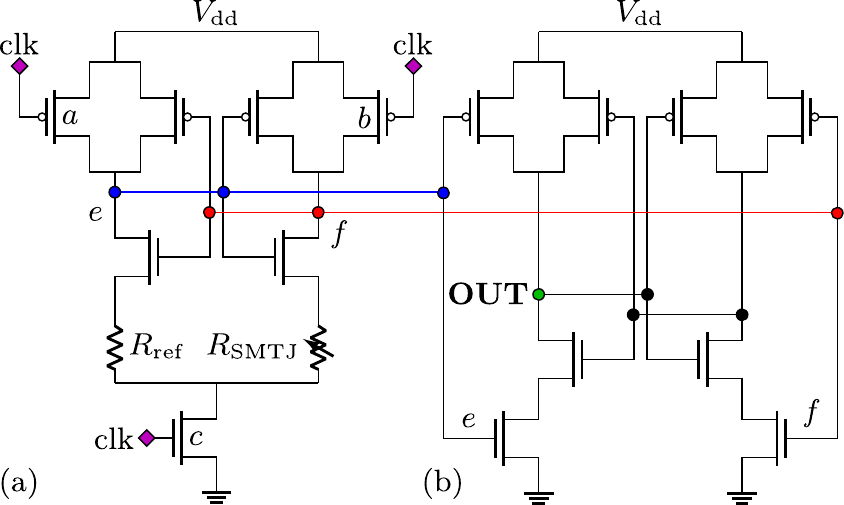}
    \caption{SMTJ readout. (a) The PCSA circuit for reading an SMTJ state~\cite{zhao2009high}. (b) The SC-PSCA includes an set/reset (SR) latch on top of the PCSA from (a), to prevent the pre-charging of nodes $e$ and $f$ from affecting the output duty cycle. The left and right ($e$ and $f$) branches of the latch are directedly wired to the $e$ and $f$ nodes of the PSCA. A more pedagogical version of this circuit is found in Fig.~\ref{fig:srlatch_dynamics}.}
\label{fig:pcsa-circuit}
\end{figure}

\begin{figure}
  \centering
  \includegraphics[width=\columnwidth]{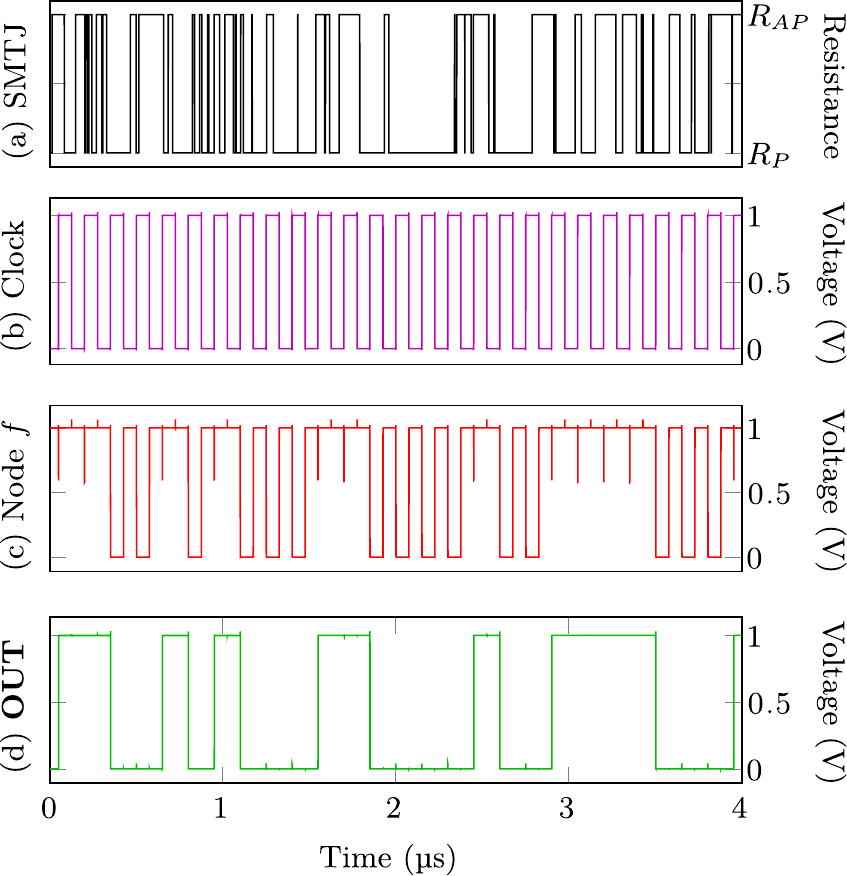}
    \caption{Circuit simulations of the SC-PCSA circuit. (a) Time-series resistance of the SMTJ due to thermal fluctuations. (b) Sampling clock (clk from Fig.~\ref{fig:pcsa-circuit}). (c) Voltage at node $f$ from Fig.~\ref{fig:pcsa-circuit}(a), the output of the standard PCSA circuit. (d) Voltage at node \textbf{OUT} from Fig.~\ref{fig:pcsa-circuit}(b), the output of the SC-PSCA.}
\label{fig:SMTJ_wf}
\end{figure}

Most previous uses of SMTJs use current biasing to vary the duty cycle of the generated bitstream, which can lead to sizeable leakage currents (typically on the order of microwatts). In large scale architectures that require many SMTJs, the resulting ohmic losses can dominate the energy consumption of the computation.  An alternate approach using a pre-charge sense amplifier (PCSA) has been proposed in the literature~\cite{zhao2009high}. In this approach, the state is read by a minimal read current pulse and is not controlled by a larger write current.  

Others have used the PCSA method successfully to generate random bits~\cite{mizrahi2018neural,PhysRevApplied.8.054045}. However, these applications only need a single random bit to be produced at a time. In stochastic computing, we need a continuous stream of random bits to be made available at the hardware level. Continuous production happens naturally in current-biased systems where the SMTJ simply sits in a voltage divider, but in the PCSA artifacts of the digital circuitry interfere with extraction of a random bitstream. 

Fig.~\ref{fig:pcsa-circuit}(a) shows the PCSA described by Ref.~\cite{zhao2009high}. It works in two cycles. When the clock signal [Fig.~\ref{fig:SMTJ_wf}(d)] is at a low voltage, the transistor at $c$ is turned off but the transistors $a$ and $b$ are on. As there is no path to ground, all wires in the circuit are brought to $V_\text{dd}$. When the clock signal goes high, a path to ground is opened at $c$, and all paths to $V_\text{dd}$ are closed. Because of the small capacitances $C$ in the transistors, there is a finite discharging time $(R_{\rm sd}+R_\text{ref})C$ in which charge drains from node $e$ to ground, and a finite time $(R_{\rm sd}+R_\text{SMTJ})C$ in which charge drains from $f$ to ground, where $R_\text{ref}$ is the resistance of the reference resistor, $R_\text{SMTJ}$ is the state-dependent resistance of the SMTJ, and $R_{\rm sd}$ is the source-drain resistance of the transistors between $e$ or $f$ and $c$. The horizontal red and blue wires provide a nonlinear interaction between these two discharging processes such that the lower resistance channel will connect to ground and the higher resistance channel to $V_\text{dd}$ after the system comes to equilibrium (Fig.~\ref{fig:srlatch_dynamics}). Note that only a small amount of charge $\propto CV_\text{dd}$ ultimately flows through the system, so ohmic losses are very small; transistor capacitances are typically on the order of 10~aF to 100~aF. Appendix~\ref{sec:srlatch} gives a more detailed discussion of the operation of the PCSA and our  proposed modification discussed immediately below.

The problem with the above process is the pre-charge phase, when the clock signal is low and the whole circuit is brought to $V_\text{dd}$. The state of the system in that phase does not represent the last measured state of the SMTJ; it is simply preparing to perform the next measurement. This can be seen in Fig.~\ref{fig:SMTJ_wf}(c). Around $t\approx2.25\,\text{{\textmu}s}$, for instance, we can see that the SMTJ is in a low resistance state [Fig.~\ref{fig:SMTJ_wf}(a)]. The output of the PCSA in Fig.~\ref{fig:SMTJ_wf}(c) nevertheless goes to a high voltage repeatedly in this timeframe. This anomalous comb structure on top of the actual SMTJ states is an artifact of the pre-charge phase. 

To address this, we attach a circuit called a set/reset (SR) latch to the PCSA design, Fig.~\ref{fig:pcsa-circuit}(b). Whenever $e$ and $f$ are different, the latch copies the values of $e$ and $f$ into its internal wires, so that \textbf{OUT} is set to $f$. When $e$ and $f$ are both brought to $V_\text{dd}$ during the pre-charge phase, the internal state of the latch is left unchanged. The behavior is explained in detail in Fig.~\ref{fig:srlatch_dynamics} in Appendix~\ref{sec:srlatch}.
The simulated state of \textbf{OUT} is shown in  Fig.~\ref{fig:SMTJ_wf}(d). With the anomalous comb structure removed, this voltage signal becomes suitable for stochastic computing applications.
 
Fig.~\ref{fig:SMTJ_wf} shows simulation results of the SC-PCSA obtained using a commercial software package and a 22~nm predictive technology model~\cite{zhao2006new,ASU_PTM}. {The simulations include modeled parasitic contributions from the transistors but not the interconnects, which would depend on layout. The interconnect capacitances would play a role at high speeds, but not the clock periods we consider. In terms of energy, the contributions from  the interconnect capacitances are negligible compared to those from the transistor capacitances.} 

The SMTJ model was implemented in Verilog-A as described in Ref.~\cite{mizrahi2015verilog} {using parallel and antiparallel resistances of 1.5~k$\Omega$ and 4.5~k$\Omega$ respectively.}  The dynamics of the SMTJ are described by the probability to switch in a given time interval, $\delta t$ of $P_{\rm switch}= 1-\exp(-\delta t/\tau)$, where $\tau$ is the mean dwell time in that state.  It is given by $\tau=\tau_0\exp(\Delta/kT)$, where $\Delta$ is the energy barrier out of the current state of the SMTJ, $T$ is the temperature, $k$ is the Boltzmann constant and $\tau_0$ is a characteristic time scale. In general, the net energy barrier depends on the applied field and voltage, {which we do not apply in this paper. These simulations use $\tau_0=10^{-9}$~s and $\Delta/kT=4$.}

\begin{figure}
  \centering
  \includegraphics[width=\columnwidth]{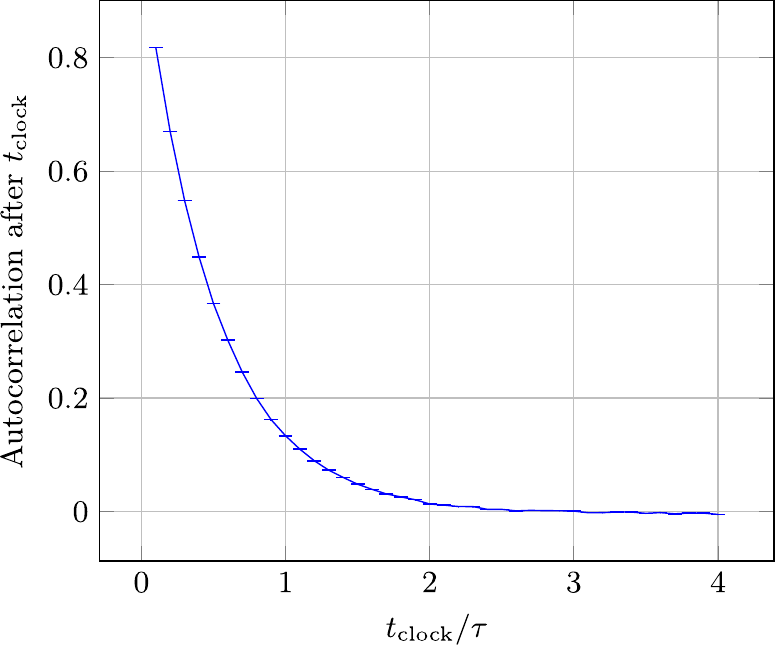}
    \caption{Autocorrelation under a lag of one clock cycle as a function of clock cycle time $t_\text{clock}$ compared to the STMJ mean dwell time $\tau$. Each point gives the mean autocorrleation time over $10^3$ trials, with each trial sampling the SMTJ for $10^3$ clock cycles. The vertical error bars give 95~\% confidence intervals on the mean.}
  \label{fig:autocorr}
\end{figure}
It is not obvious that SMTJ output sampled at a given time is decorrelated from the output sampled on the same device a single clock cycle later. If the SMTJ is read too frequently, the state of the device has no time to change, and the generated bitstream will have strong autocorrelation. This autocorrelation can suppressed by increasing the sampling interval. In Fig.~\ref{fig:autocorr}, we see from numerical simulation that the autocorrelation between two time-adjacent samplings vanishes once the clock interval is chosen to be more than twice the mean dwell time $\tau$ of the SMTJ. In other words, the clock cycle should be chosen so that the expected ``period'' $2\tau$ of a $P\rightarrow AP\rightarrow P$ cycle fits within a single clock cycle. When the SMTJ switches more slowly than the clock, the autocorrelation increases linearly with $\tau$.


Establishing an appropriately large clock cycle is crucial. In many stochastic computing applications, autocorrelation can introduce not only energy inefficiency but functional incorrectness. In the application we present in Sec.~\ref{sec:neuralnetworks}, for instance, we will purposefully delay time signals from each other in order to suppress cross-correlations among them. This only works if the initial signals themselves have vanishing autocorrelation.

If the continued development of SMTJ technology could engineer a sufficiently small autocorrelation time for the thermally-induced magnetic dynamics---that is, a speed comparable to the switching speed of the CMOS gates---then the integrated ohmic power loss will be similar to losses in the CMOS itsef, and it may then become reasonable to place a \emph{static} read current across the devices, as proposed in Refs.~\cite{roy2018perspective,ganguly2017reservoir,PhysRevX.7.031014}. This would provide a continuous-time random telegraph signal, opening up the possibility of running asynchronous computations. We compare the energetic performance of such an approach to our proposal in Sec.~\ref{sec:bitstream-gen}, in the context of current SMTJ technology. We explore this limit in detail in Appendix~\ref{app:p-bits}.

Since the SC-PCSA is charge based, the dwell time of the SMTJ does not change the energy expenditure of the circuit (so long as the mean dwell times are not faster than the equilibration time of the PCSA, about 1~ns). Ideally, then, we would like the SMTJs to fluctuate as quickly as possible; in general, we would also hope for uniformity of dwell times and operation ranges. Measured dwell times range between 1~{\textmu}s and 0.1~s depending on the operating regime~\cite{rippard2011thermal}.  The fluctuation rates are highly sensitive to applied fields and currents.    The two state fluctuator model used in this paper is not valid for fluctuator frequencies approaching the $10^{-9}$~s time scale of magnetization reversal~\cite{faria2017low}, but for applications like ours in which a reference resistor is present, Kaiser et al.~\cite{kaiser2019ultrafast} show that frequency scales can exceed the 1~GHz regime.  The achievable time scale depends on the differences in the resistance, proportional to the tunneling magnetoresistance of the magnetic junction, that can be detected and the size of the current needed to do so. 

The resistance and magnetoresistance of the tunnel junction are the same as those developed for memory applications and so should have margins sufficient for effective circuit design \cite{tsuchida201064mb}.  In memory applications, however, the state of device is switched by the spin transfer torque (STT) that is generated by the current passing through the device. Substantial work has been done to make that switching current as low as possible \cite{jiang2004substantial}.  {In the present application, by contrast, we want the current response to be as weak as possible, so that the read current has a minimal effect on switching rates.  Unfortunately, the current needed to influence the switching behavior apparently \cite{sato2017magnetic} scales to smaller values as devices become smaller. Increasing the this current  requires additional research; recent work suggests that devices with easy-plane anisotropy may be preferable, in this sense, compared to perpendicularly magnetized devices~\cite{hassan2019low}}.

There has been considerable work in the recent literature to use MTJs or SMTJs as the fundamental units for truly random number generators~\cite{fukushima2014spin,PhysRevApplied.8.054045,parks2018superparamagnetic,choi2014magnetic}. For general purpose applications, random number generators are required to pass certain tests of randomness; the NIST statistical test suite \cite{rukhin2001statistical} is usually taken as a standard benchmark for validating good randomness in that sense. {An important criterion required by that test suite is that the mean value of a bitstream has a long time average of 50~\%. To date, all MTJ-based solutions that pass those tests require XOR'ing eight devices together to eliminate bias.  Given SMTJs with similar enough dwell times in each state as we assume here, we expect that we could combine SC-PCSAs in a similar manner and pass the statistical tests needed for random number generation.}

{In this work, we do not subject our SC-PCSA to these randomness tests, as they are not necessary for our application space. Stochastic computing aims to encode values in the expected value of random bitstreams.  It is traditionally implemented with low bit-resolution pseudorandom number generators that have poor randomness properties; some stochastic computing has even been done with entirely deterministic bitstreams~\cite{jenson2016deterministic,najafi2018deterministic}.} The important properties needed for stochastic computing are simply small enough cross-correlation between different devices and small enough autocorrelation after some time has passed.  The correct metric for the success of our generator is {not that it can produce unbiased random bits for cryptographic or scientific applications, but} that it can be used to carry out stochastic computing calculations. We show that it can in a complex neural network architecture in Sec.~\ref{sec:neuralnetworks}.

\section{SMTJ programmable bitstream generator}

\label{sec:bitstream-gen}

In order to generate arbitrary bitstreams with $n$-bit precision, we develop a composite cell that can be programmed based on values stored in static random access memory (SRAM) cells. If the programmed value is $1/2$, we need only tap the output of one of the SC-PCSA cells discussed in Sec.~\ref{sec:smtj-pcsa-readout}. Otherwise, we perform a sort of binary search on the unit interval. Given a stochastic signal carrying the probability value $k/2^{n-1}$, it is synchronously fed into a \texttt{NAND} gate together with a $1/2$ probability signal from a new SC-PCSA cell to produce a signal with probability $1-k/2^n$. A simple multiplexer (MUX) can then optionally route the signal through an inverter, allowing us to access values of both $1-k/2^n$ and $k/2^n$.

\begin{figure}
    \centering
    \includegraphics[width=\columnwidth]{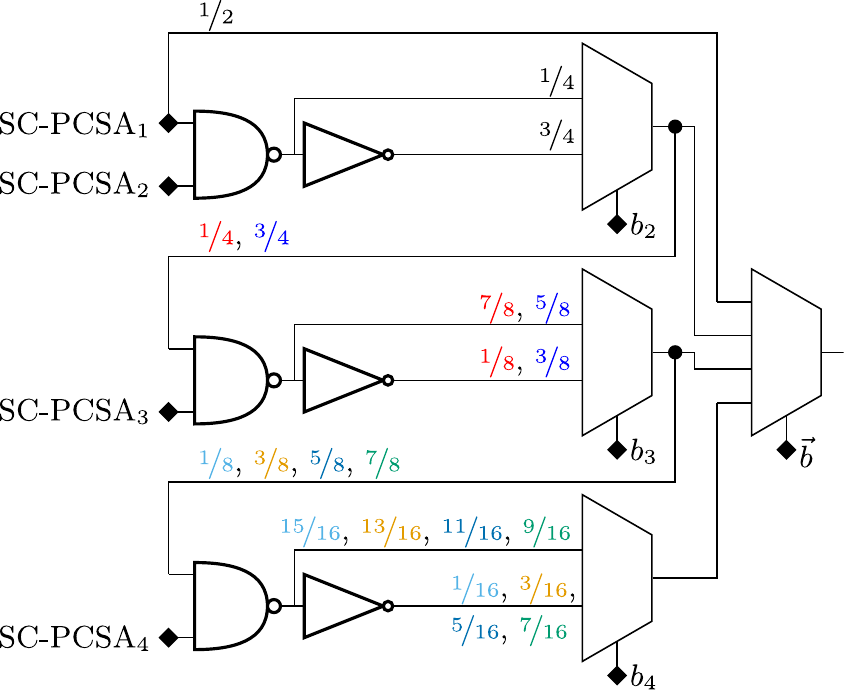}
    \caption{Bitstream generator with four programmable bits. Each row is a recursive subdivision unit. Assuming input probabilities of $1/2$ from the SC-PCSAs, the topmost 2-input multiplexer outputs $1/4$ or $3/4$ depending on $b_2$; the middle multiplexer $1/8$, $3/8$, $5/8$, and $7/8$, depending on $b_2$ and $b_3$; and so on. SRAM control-and-decode circuits, which set the bits $\vec b$ appropriately for the desired output probability, are hidden for clarity.}
\label{fig:SMTJ_prog_gen}
\end{figure}

Iterating this recursive subdivision operation allows us to access all multiples of $1/2^n$ using $n$ SC-PCSA cells. Values for zero and unity can be trivially implemented on the end of the circuit by replacing the stochastic signal with a constant voltage. In our simulations of a neural network later in this paper, we restrict ourselves to the 4-bit case, as illustrated in Fig.~\ref{fig:SMTJ_prog_gen}, allowing our synaptic weights to express integer multiples of $1/16$.

Though strict tests of randomness are not necessary for our purposes, careful attention to correlation is. Statistical correlations can arise in many ways in a stochastic computing circuit. We distinguish between two types: \emph{graph correlation}, and \emph{source correlation}. To understand the former, suppose a circuit designer makes a simplifying assumption that the inputs to a particular circuit node represent statistically independent stochastic processes. This may not be true, depending on whether the computations producing those inputs overlapped at some point earlier in the circuit. If they are not statistically independent, the circuit output statistics will differ from the value expected by the designer. We call the correlations leading to such errors graph correlations, as they arise due to reconvergent fanout in the computational graph of the circuit. One method for alleviating these correlations is called isolation, and has recently been discussed in Ref.~\cite{Ting_2016}. We make use of these and discuss their implementation and impact in Sec.~\ref{sec:neuralnetworks}.

The programmable bitstream generator described above addresses a more insidious form of correlation that can arise when the original bitstreams are not random. In typical implementations of stochastic computing, the LFSRs and other commonly used sources of pseudorandom bitstreams are periodic. Overusing the same LFSR design at multiple instances in a circuit can lead to correlation-induced errors that are difficult to predict~\cite{Salam1994}. We call these source correlations; they are induced by inconsistent assumptions about bitstream generators on the boundary of the computational graph.  One of our motivations for SMTJ-based stochastic computing is to solve the problem of source correlations. This is especially important in neural networks, where hundreds or thousands of inputs can all fan-in to the same circuit node. To avoid source correlations in this case, pseudorandom circuit designs become prohibitively energy-hungry, but our SMTJ-based approach scales to any degree of fan-in with no source correlation present.

\begin{figure}
\centering
\includegraphics[width=\columnwidth]{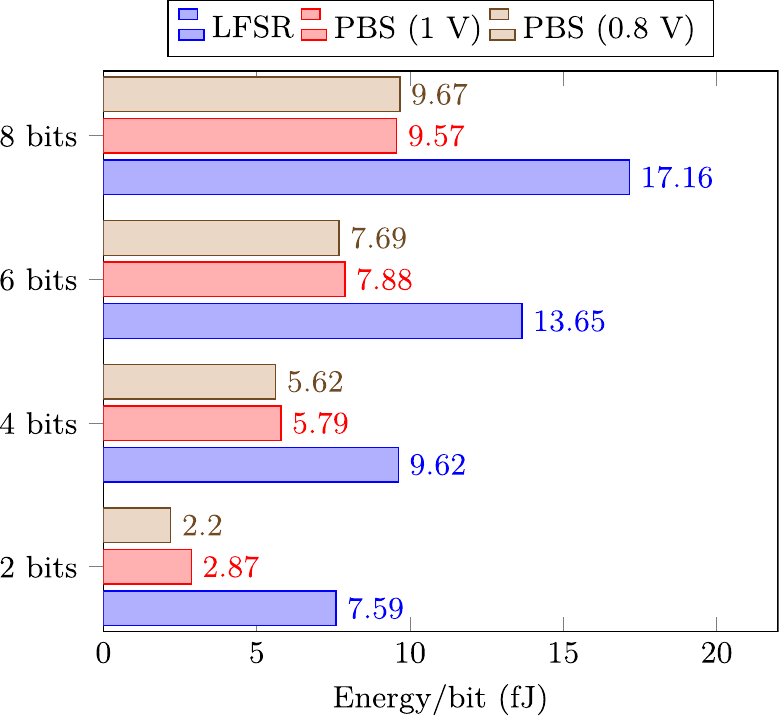}
\caption{Energy efficiency comparison between our programmable bitstream generator (PBS) at supply voltages of 1~V and 0.8~V and a traditional LFSR with binary comparator. Plotted horizontally is the energy needed to produce a single new element of the output stochastic bitstream. The $N$-bit PBS uses $N$ SMTJs where the $N$-bit LFSR uses an $N$-bit register, ensuring a fair equiprecision comparison.}
\label{fig:energy-nums}
\end{figure}

Energy numbers for our approach at various bit-precisions and power supply voltages are shown in Fig.~\ref{fig:energy-nums}. Though lower supply voltages should generally lead to lower energies, we see that for $n = 8$ the energy per cycle is actually higher for the lower supply voltage. In order to get high reliability operation of the SC-PCSA at low supply voltage, the drive strength of some transistors needs to be increased. This results in the 0.8~V SC-PCSA requiring more energy than its 1~V counterpart. As the number of SC-PCSAs in the readout circuit increases at high bit-precision, this anomalous energy cost at low supply voltage can overwhelm the nominal savings in the rest of the logic tree.

Fig.~\ref{fig:energy-nums} also compares SC-PSCA energy to the energy consumption of an LFSR with the same bit precision. For a fair comparison, we designed the (Galois type) LFSRs according to standard maximum sequence length designs, and using the same technology node and predictive technology models as for the SC-PSCA simulations. We found that the SC-PSCA performed about twice as efficiently across a range of bit precisions. This result is predictable from the transistor count of the two designs; the SC-PSCA has 15 transistors, whereas a clocked, \texttt{NAND}-gate based digital flip flop (the fundamental unit of the LFSR) has 20. The circuitry of comparator in the LFSR-based is also slightly larger than the \texttt{AND}-{NOT}-{MUX} rows of Fig.~\ref{fig:SMTJ_prog_gen}. The LFSR-based scheme additionally contains some small number of \texttt{XOR} gates, though the number of \texttt{XOR}s is an algebraic property that does not necessarily scale with the number of bits, $N$.

We note that several proposals address the cost of LFSR-based pseudorandom number generator by extracting combinatorial subsets of the bits in a large LFSR~\cite{ichihara2014compact,kim-2016}. Though these shared-LFSR methods amortize the energy cost by sharing it over many pseudorandom bits, the resulting correlation between these bitstreams is $1.5$ to $2$ times higher than the isolated LFSR case~\cite{kim-2016}. In correlation-sensitive applications, this is clearly disadvantageous. In applications that have been engineered to be correlation insensitive, one could imagine applying the same shared generator techniques to our programmable bitstream generator; the same energy/correlation trade-off should apply. We leave the details of such a device to future research.

Several related configurations for using magnetic tunnel junctions have been proposed that could be used for generating stochastic bitstreams.  Most fall into two broad categories, those based on nominally stable magnetic tunnel junctions that are brought into an unstable state by a current \cite{fukushima2014spin,choi2014magnetic,mondal2019energy} and those based on SMTJs that are current biased to control the expected value of the bitstream \cite{faria2018implementing,jia2018spintronics,sengupta2016probabilistic,PhysRevX.7.031014,ganguly2017reservoir,hassan2019voltage, shim2017ising}.  The write currents in both of these approaches lead to significant ohmic losses, which are disadvantageous for this stochastic computing.  We note however that in many cases, these configurations have been proposed for  applications for which the ohmic losses may not be quite as important.

We estimate the ohmic losses for pulsed destabilizing of stable MTJs in Appendix~\ref{app:switchedMTJs}. The circuits reported in Ref.~\cite{choi2014magnetic} are based on MTJs with average resistance $R= 1000~\Omega$, and use an average write voltage of $217.5~\text{mV}$ for $8.75~\text{ns}$. The ohmic losses are then 414~fJ per bit. Ref.~\cite{mondal2019energy} uses similar techniques on MTJs specifically tailored to the generation of stochastic computing bitstreams (the same task we consider here); they report average costs of 526~fJ per bit. Both estimates are significantly higher than the 10~fJ estimated for the SC-PCSA-based approach. 

To estimate the ohmic losses (see Appendix~\ref{app:p-bits} for details) for $p$-bit style current-biased SMTJs, we assume a supply voltage of $V_\text{dd}=$~1~V and favorable resistances of $R_{AP} = 100~\text{k}\Omega$ and $R_P = 50~\text{k}\Omega$, and find ohmic losses of about 500~fJ per bit for the 150~ns clock cycle we consider; {the clock period is set by the autocorrelation time of the SMTJ}.  In the SC-PCSA-based approach described above, the energy per bit of 10~fJ for the whole circuit is roughly independent of the clock cycle.  {Note that}, for the current controlled approach, the energy per bit decreases linearly as the clock cycle decreases, decreasing the advantage of the SC-PCSA-based approach.

We cannot expect the manufacturing margins of real SMTJ devices to be ideal.  The fabrication of magnetic tunnel junctions has been optimized in commercial fabrication facilities~\cite{chung20164gbit,rho201723} in the nonvolatile regime that is desirable for memory applications.  SMTJs have only been studied for larger devices fabricated in laboratory settings~\cite{rippard2011thermal}.

Real devices will have distributions of all of their properties.  Three important ones are variations in barrier heights (and hence characteristic dwell times), variations in biasing around the perfect $p=1/2$ {that we have assumed above, and variations in the device resistances.} The insensitivity to a distribution of barrier heights in addressed in Fig.~\ref{fig:autocorr}. Essentially uncorrelated bitstreams are generated as long as the mean dwell times are shorter than the clock cycle and longer than some lower limit set by the response time of the PCSA, on the order of 1~ns for the circuits considered here.  Assuming a prefactor of $\tau_0\approx 1$~ns in the transition time distribution, this approach is insensitive to fluctuations in the barrier height for $0 < \Delta/kT < 5$. Since considerable effort has been required to keep energy barriers large as MTJ devices become smaller \cite{perrissin2018highly,liu2018top}, it seems that it should not be difficult to fabricate devices with small enough barriers to enable thermally driven transitions.  Variations in the barrier height should not pose a difficulty if it is possible to maintain the roughly 10~\% variation in barrier heights~\cite{tsunoda2014area} as devices are scaled down.

\begin{figure}
  \centering
  \includegraphics[width=\columnwidth]{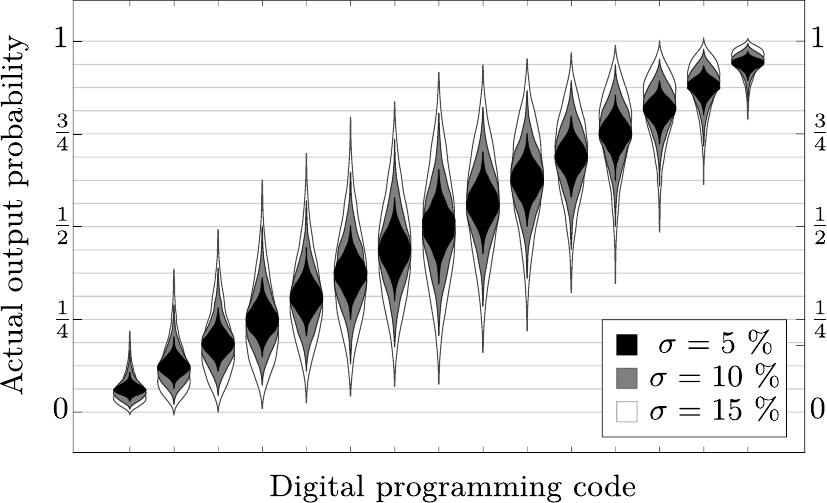}
  \caption{Output probability distributions for the four bit generator induced
    by imprecision in SMTJ fabrication. The three different standard deviations
    correspond to the three distributions in Fig.~\ref{fig:beta-dists}. The
    width of a bubble at a particular vertical coordinate gives the
    probability density that the generator outputs that probability in its
    bitstream. Note that the three, two, and one-bit versions can be extracted by simply restricting this plot to outputs with mean values that are multiples of $1/8$, $1/4$, and $1/2$, respectively.}
  \label{fig:distribution-plots}
\end{figure}

Having an expected value of 0.5 for the output of the bitstream generator requires that the energies of the parallel and antiparallel states are the same.  Such equality in turn requires that the fringing fields acting on the free layer be close to zero.  It is difficult to fabricate devices with precisely these fields and there will naturally be a distribution of relative energies and hence dwell times.
Fig.~\ref{fig:distribution-plots} gives the effect of the variation in the relative dwell times on the output of a four-bit stochastic bitstream generator. We assume that the fluctuations in the relative dwell times in the parallel and antiparallel states are distributed around zero with standard deviations given in the figure (see Appendix~\ref{app:variability} for details).  These distributions should be compared to the expected distribution for perfectly balanced bitstreams that are sampled for a finite sample size.  For the results presented below based on using bitstreams of length 128, the expected distributions of values found from perfectly balanced bitstreams are similar to the distributions found for 10~\% variations in the distribution of the relative dwell times.  

{Finally, we address whether the distributions of  resistances would adversely affect the performance of our bitstream generator. We have avoided the use of spintronic control over the random telegraph noise, so resistance variations do not directly affect the probability of our bitstreams. The main failure mode would arise from resistance variations, that is, if the reference resistor $R_\text{ref}$, selected before devices are grown and tested, were to not fall between the resistances $R_\text{P}$ and $R_\text{AP}$ of the parallel and antiparallel states.}

{Recent data from fabrication facilities~\cite{park2018low,xue2018process} show that the the manufacturing margins should be tight enough to accomplish this.  The 3$\sigma$ variability in the parallel state resistance is under 10~\%. The $3\sigma$ variability in the tunneling magnetoresistance is only 2~\% to 3~\%. If we target a $1.5~\text{k}\Omega$ resistance for $R_\text{P}$, and a $4.5~\text{k}\Omega$ resistance for $R_\text{AP}$, then a reference resistance of $3~\text{k}\Omega$ is about 15$\sigma$ away from both resistances, providing an adequate manufacturing margin even if the magnetoresistance were much smaller than the assumed 200~\%. In addition, neural network applications, which we have considered in some detail in the paper, can be robust to single-device failures, if the failure of particular devices can be detected (which it is easy to imagine doing here).}

{The data we cite above is for high-barrier magnetic tunnel junctions developed for magnetic random access memory (MRAM) but manufacturers are pushing toward the superparamagnetic regime; recently announced~\cite{alzate2019arrayLevel} work on embedded MRAM devices works with retention times on the order of seconds to milliseconds. This is not too far from the SMTJ speeds fabricated in academic labs, and the reported fabrication distributions continue to be satisfactory in this lower barrier regime. Based on these reports, our proposed approach should be resistant to device fluctuations within the range of fluctuations that can be expected from a dedicated fabrication process.}

\section{Application to neural networks}
\label{sec:neuralnetworks}

Artificial neural networks, initially inspired by the structure of the brain \cite{mcculloch1943logical}, have become one of the most powerful classes of algorithms in machine learning. 
These neural networks are composed of two fundamental units: neurons, which sum inputs and apply a nonlinear activation function to the resulting sum; and synapses, which multiply the output of one neuron by a real-valued weight and pass the result to a downstream neuron. Traditionally, the numerical values that propagate through neural networks correspond to rates at which neural spikes propagate through an actual biological brain.

The origins of stochastic computing lie in the observation that time series data of stochastic spike trains in the brain could be modeled by stochastic jumps from ground to $V_\text{dd}$ in a logic circuit~\cite{von1956probabilistic,poppelbaum1967stochastic,gaines2019origins}. It is no surprise, then, that neural network structures have been implemented successfully and energy efficiently in recent stochastic computing work~\cite{Ren_2017,Li_2018,ardakani2017vlsi,Onizawa2019}. Rather than carrying out high level arithmetic and logic operations to ``theoretically predict'' a neural network's output, stochastic computing implements neuromorphic models of the network in CMOS circuitry. The network operation is ``experimentally simulated'' by the physics of the circuit, and the results are obtained by monitoring the time series voltages at the network output.

A crucial synergy involved in this scheme is the inherent parallelism of both neural networks and stochastic computing. Whereas the many mathematical operations involved in a neural network layer would need to be computed serially in a traditional computing environment, the physical nature of the stochastic computer means that these operations are all run simultaneously.  There is a sizeable body of recent work that uses this principle to build efficient, stochastic-computing-based neural networks. The first stochastic computing implementation of a deep convolutional neural network, a particularly important neural network topology with broad applications to image processing, was proposed by Ren \emph{et alia}~\cite{Ren_2017} and further optimized in Refs.~\cite{Li_2018,li2017hardware}. These works draw heavily on new ideas in the stochastic computing literature, including massively parallel generation of pseudorandom bitstreams~\cite{kim2016energy}, state-machine based nonlinear activation functions~\cite{Kim:2016:DET:2897937.2898011,ardakani2017vlsi}, and aggressive use of correlation insensitivity~\cite{alaghi2013exploiting}.

A common approach in the stochastic computing neural network literature is to construct a neuron unit from multiplexer-based addition composed with a state-machine-based nonlinearity. These operations are costly in terms of CMOS transistor counts. We take a simpler approach: our entire neuron is a single logical \texttt{OR} gate. 

{In our approach, a single multi-input logical \texttt{OR} gate simultaneously, approximately, and inseparably performs both the summation and non-linear activation.} Though for small input magnitudes an \texttt{OR} gate performs addition of probabilities $p_1 + p_2 + O(p^2)$, nonlinear corrections become important as the input probabilities increase. Utilizing this property to our advantage, we harness that nonlinearity directly as our neuron's nonlinear activation function. This approach was originally proposed in ~\cite{Tomlinson_1990,Kim_Shanblatt_92,Young_Chul_Kim_1995_random_noise,Young_Chul_Kim_1995_architecture}, but until now has been unable to scale beyond small networks due to correlations between LFSRs. Our use of truly random bitstreams sourced from SMTJs opens the possibility of using this very small and efficient neuron in modern-scale neural networks. The synapses in our work are \texttt{AND} gates, which naturally implement multiplication on probabilities, with inputs from the output from the previous neuron and from a fresh stochastic bitstream encoding the weight (in our case, provided by an SMTJ programmble bitstream generator).
 
\begin{figure}
  \centering
  \includegraphics[width=\columnwidth]{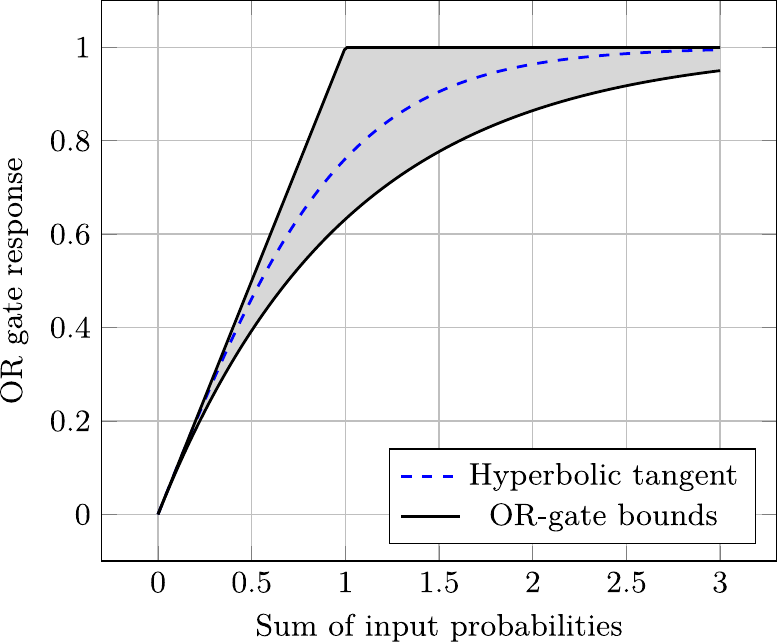}
  \caption{Activation pseudo-function of the \texttt{OR} gate in terms of its
    bounds. The shaded region indicates possible output values in the large fan-in case. For comparison, the dashed line plots the hyperbolic tangent function, a common neural activation in deep learning applications. Like the hyperbolic tangent, the \texttt{OR} gate operation is asymptotically linear near the origin and saturates exponentially to unity.
    }
  \label{fig:or-gate-activation}
\end{figure}

The probabilistic response of an \texttt{OR} gate in stochastic computing is analogous to its Boolean response; the output probability is given by
\begin{equation}
    \label{eq:or-gate}
    P_{\texttt{OR}}(\bm p) = 1 - \prod_{j=1}^N (1-p_j),
\end{equation}
where $\bm p$ is the vector of input probabilities. Eq.~\eqref{eq:or-gate} holds when each input probability $p_j$ is an independent random variable. Note that this condition generally fails in stochastic neural networks that use LFSR-based bitstream generation, especially when the number of inputs to a neuron approaches or exceeds the periodicity of the LFSR design being used.

Unlike a traditional artificial neuron, the \texttt{OR} gate's action on probabilities cannot be factored into two separate processes of summation and activation. But by examining its limiting cases, we can see that the \texttt{OR} gate does \emph{approximately} perform summation-and-activation functionality: for the large fan-in case, where $N > 10$,  the lower bound of the OR gate output approaches $1-\exp(-\sum_jp_j)$. We plot this bound, and all of the other allowed output values, as a function of the input probability sum in Fig.~\ref{fig:or-gate-activation}. The functional form of this distribution is similar to the hyperbolic tangent function, a widely used activation function in machine learning.

The inputs to a neuron are multiplied by real numbers often called synaptic weights. We represent each weight using a programmable bitstream generator, and each multiplication using an AND gate.  The \texttt{OR}-gate neuron, together with \texttt{AND}-gate synapses, form the fundamental unit of our proposed neural network architecture. This is an extremely energy efficient primitive cell, and we have to accept multiple constraints in order to use it. The most striking constraint is that, since all values in our neural network are represented by probabilities between zero and one, our network nominally lacks any form of inhibition (classically implemented with negative numbers) {as reflected in the fact that the activation function in Fig.~\ref{fig:or-gate-activation} is not defined left of the origin.} 

{Without making changes at the algorithmic level to accommodate the lack of inhibition, the \texttt{OR}-gate neuron would be largely useless. To facilitate the use of this neuron, we replace the standard weight matrices by two sets of weight matrices, one for inhibition, $\boldmath{w}^I$, and one for excitation, $\boldmath{w}^E$, that are merged to provide a form of inhibitory activation for the network. The inhibitory subneuron uses a \texttt{NOR} gate rather than an \texttt{OR} gate so that large inhibitory input supress the neuron output. The binary output $y\in\{\mathtt{0},\mathtt{1}\}$ of each neuron is given by 
\begin{equation}
       y = \underset{\mathtt{OR}}{\underbrace{\left[1-\prod_j(1-w_j^Ex_j)\right]}}\times\underset{\mathtt{NOR}}{\underbrace{\left[\prod_j(1-w^I_jx_j)\right]}},
       \label{eq:or-and-nor}
\end{equation}
where the outermost multiplication is implemented with a single \texttt{AND} gate. The use of both excitatory and inhibitory subnetworks introduces a higher dimensional symmetry, which the network can break to accomplish dynamical inhibition (or excitation) and successfully learn correct weights for classification. More details are discussed in Appendix~\ref{sec:dual-architecture}}.

{Since the \texttt{OR} gate neuron is unusual, the network training algorithms must adapted to it.} We derive the backpropagation equations for this neuron and the inhibitory/excitatory subnetworks for use in training, and adapt standard training algorithms to accommodate the hard constraints on allowed weights. The details of this codesign process are elaborated upon in Appendix~\ref{app:nn-details}. {One important note is that Eq.~\eqref{eq:or-and-nor} prescribes the correct boolean output at a given clock cycle, but is formally inaccurate as a prescription for the expectation value of $y$ over many cycles. It treats the inputs $\bm x$ to the \texttt{OR} and \texttt{NOR} gates as independent random variables, when in fact they are identical and thus perfectly dependent on each other. Nevertheless, we found empirically that the use of Eq.~\eqref{eq:or-and-nor} for network training performed well enough for our purposes. Refining the training process to use a more accurate expression could improve network performance.}

To demonstrate the effectiveness of our stochastic neuron and synapses, we implement {a stochastic approximation of LeNet5~\cite{lecun1998gradient}, a convolutional neural network. Formally, LeNet5 is based on floating point arithmetic and specific choices of neurons. We instead use the \texttt{AND} and \texttt{OR} gates for synaptic and neural operations, as outlined above, in a stochastic computing framework. We also make simple changes to the layer structure to account for our restriction to nonnegative probabilities, which are outlined in Appendix~\ref{app:nn-details}. This benchmark task allows for an easy comparison to existing literature. Ref.~\cite{Li_2018} treats the standard form of LeNet5 (using stochastic arithmetic) as} a fixed boundary condition at the top of the stack, and optimizes a scaffolding of stochastic computing hardware around this software-oriented neural network design. The implementation of LeNet5 we present here is instead based on the kind of cross-stack reasoning we just outlined: we relax the strict definition of LeNet5 as a boundary condition and use the resulting flexibility to accommodate our energy efficient choice of neurons and synapses at lower levels. 

\section{Evaluation and Results}
\label{sec:eval}

To evaluate our implementation of LeNet5, we train and test our neural network on the MNIST dataset. The MNIST dataset comprises 60000 training images and 10000 test images of $28\times28$ pixel images. First, we perform offline training on an analytic model based on probabilities rather than stochastic representations of probabilities, that is, on a traditional software model using the unusual topology, constraints, and activation functions of our proposed network. Though the network does train, we find the process to be significantly noisier than training of classical LeNet5 models under the same hyperparameters. We speculate that the noise arises from the unfactorability of the summation-nonlinearity approximated by Eq.~\ref{eq:or-gate}, as well as the hard constraint that all weights must be between zero and one. 

Among successfully trained networks, we find that we frequently achieve around 98~\% test set accuracy with the analytic model applied to MNIST,  within 1~\% of the 98.9~\% accuracy achieved by the original LeNet5 authors~\cite{lecun1998gradient}. 
We then discretize the trained weights and biases by rounding them to the nearest multiple of $1/16$. These are loaded into a stochastic simulation of the logical architecture, complete with stochastic processes to generate the SMTJ statistics. 

\label{sec:isolator-perf}
To deal with residual graph correlations induced by reconvergent fanout of the signals into and out of each neural network layer, we insert {what we call} an isolator mask---an array of randomly chosen (but static) pixelwise temporal delays---after each network layer.  We discuss in more detail in Appendix~\ref{sec:decorr-arch} how isolators can be used to achieve decorrelating effects by shifting signals relative to each other in time. Each pixelwise, integer-valued delay is chosen uniformly at random from the interval $[0,\delta]$, where $\delta$ is the maximum delay. In our LeNet5 architecture, this leads to a total of $L=6$ decorrelation layers. The maximally delayed route from input to output, then, has delay length $W=L\delta$, which is the amount of time we must wait before collecting $N$ usable data points at the output. 


\begin{table}[]
\centering
\caption{Energy breakdown on worst-case analysis of LeNet5 case study for an SMTJ-driven computation. The rightmost column corresponds to the $N=128$, $\delta=16$ network configuration.}
\begin{tabular}{@{}lrrr@{}}
  \toprule & $\frac{\text{Energy/fJ}}{\text{Element cycle}} $ & $\frac{\text{Energy/nJ}}{\text{LeNet5 cycle}}$ & $\frac{\text{Energy/nJ}}{\text{inference, at 97~\% accuracy}}$\\ \midrule
  $V_\text{dd} = 1\;\text{V}$\\ \addlinespace
  Weights & 5.80 & 0.50 & 112.80 \\
  Neurons & 3.11 & 0.04 & 8.09 \\
  Isolators & 19.75 & 0.11 & 25.72 \\ \cmidrule{3-4}
  \textbf{Total (nJ)} &     & \textbf{0.65} & \textbf{146.62}\\ \midrule
  $V_\text{dd} = 0.8\;\text{V}$\\ \addlinespace
  Weights & 5.62 & 0.49 & 109.39 \\
  Neurons & 2.47 & 0.03 & 6.44 \\
  Isolators & 15.02 & 0.09 & 19.56 \\ \cmidrule{3-4}
  \textbf{Total (nJ)} &     & \textbf{0.60} & \textbf{135.40} \\
  \bottomrule
\end{tabular}
\label{tab:energy-nums}
\end{table}

\begin{figure}
\centering
\includegraphics[width=\columnwidth]{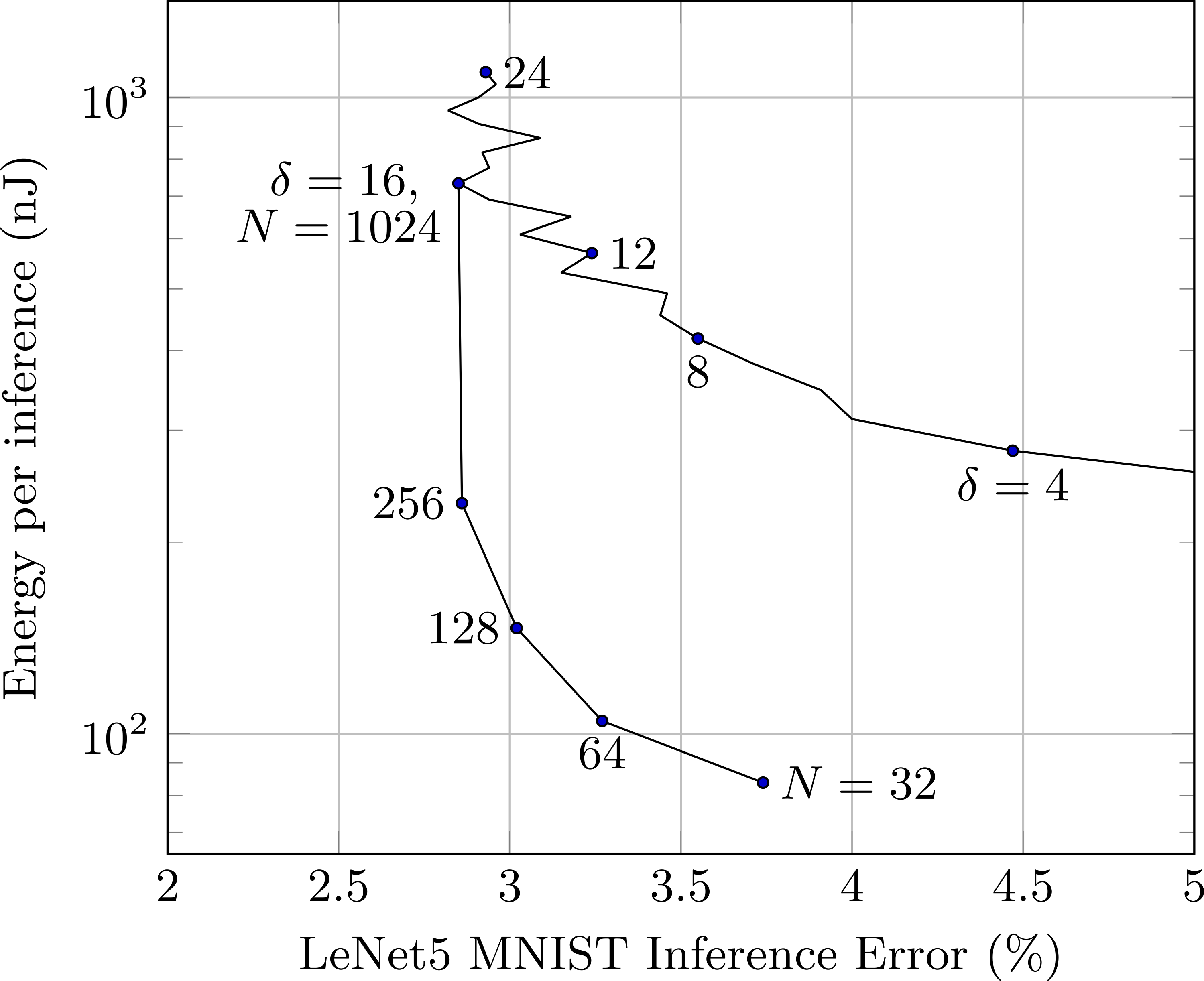}
\caption{{Evaluation of our network performance for various hyperparameters. The upper curve shows variation in $\delta$, the maximum number of isolators decorrelating each of the $L$ hidden layers. This affects the energy by increasing the total warm-up time $L\delta$ needed before network output becomes meaningful. The lower curve tracks $N$, the number of meaningful data points collected \emph{after} the warm-up period concludes.}}
\label{fig:n-delta-curves}
\end{figure}

In the {upper black} curve on Fig.~\ref{fig:n-delta-curves}, we plot the performance of a trained network on the test datasets as a function of the maximum delay $\delta$ and length of collected bitstream $N$. Performance improves exponentially with increasing $\delta$, eventually saturating around a maximum delay length of 16---and, therefore, a mean delay length of 8. We also find{, in the lower curve,} that increasing $N$ provided diminishing returns beyond $N\approx 128$. We use this $(\delta,N)=(16,128)$ configuration to report the rest of the numbers in our paper. The fluctuations that become evident in the plot at high $\delta$ arise from the pseudorandomness of the isolator mask configuration. A systematic approach would search through multiple mask configurations until an appropriate one is found for the given trained network.

\label{accuEnergy}
To compute the energy efficiency of our proposed architecture, we extract the mean activity factors from weights and neurons in the network over an inference pass on the MNIST test data set. As our simulation is purely logical (that is, it works at the level of clocked bits rather than dynamical voltages), we find it unwieldy to extract the expected activity factors from inside the circuit-level composite gates of the \texttt{OR} gate neuron. Instead, we convert the activity factors from the logical model into the \emph{worst case} activity factors for the realistic architecture. 

To produce our worst-case analysis, we assume that \emph{every} switch in the voltage signal causes a maximal number of switches in the \texttt{AND} gate synapse \emph{and} the downstream \texttt{OR} gate neuron. We apply this assumption to the activity factors found in our stochastic model, together with circuit power estimates based on predictive technology models~\cite{zhao2006new,ASU_PTM}. The results for the $N=128$, $\delta=16$ case are listed in Table~\ref{tab:energy-nums}. For $V_\text{dd} = 1\,\text{V}$ at the 22~nm node, CMOS switching events in \texttt{OR} gate neurons and \texttt{AND} gate synapses together account for 8.09~nJ. 

Random bitstream generation and isolator-based decorrelation are more expensive than the computational graph. For these estimates we again use worst-case activity factors. In the full LeNet5 architecture, the pre-charge sense amplifier circuits generating network weights and inputs account for about 113~nJ for a full inference. The programmable isolator buffers are each relatively expensive, but fewer were needed compared to the weights.   The total cost for the isolator circuits is around 26~nJ. 

We also test the behavior of our architecture at a reduced supply voltage. For voltages below 0.7~V, the PCSA performance begins to degrade, causing errors in the output bitstream. At 0.8~V, the total energy consumed by the entire architecture drops from 147~nJ to 135~nJ. These energy estimates are also displayed in Table~\ref{tab:energy-nums}.

One likely source of network inference error that remains after accounting for latency and isolator length is the quantization of weights from the analytic model to discrete values allowed by the bitstream generator. Co-design of hardware with deep neural networks has made the study of low-precision neural networks a topic of considerable recent interest~\cite{sze2017efficient}; quantization-aware training methods are now being developed both for pure CMOS systems~\cite{gupta2015deep,zhou2016dorefa,hubara2017quantized} and platforms with nanodevice integration~\cite{song2017quantization,yang2018quantized}. Applied to systems like ours, these may offer the possibility of improved inference performance at fewer and fewer bits, improving energy consumption and network latency. 

With that said, the robustness of our stochastic model after rounding the weights of our analytic model expresses a robustness to error characteristic of neural networks. We speculate that, since network performance was robust with respect to rounding weights to multiples of $1/16$, it should be equally robust had the weights been rounded to values slightly different than the multiples of $1/16$. If that is indeed the case, then systems like ours would enjoy a degree of robustness against variation of the type shown in Fig.~\ref{fig:distribution-plots}.

Although some designs for modern neural networks can work at four bit (or sometimes lower) precision, it does not necessarily follow that one could successfully use an $N=16$ stochastic computing network. Stochastic computers have two types of precision that are intrinsically linked: representational precision, and sample precision. Collecting $N$ time steps means that $N$ different values are representable on the output, but it also means that our certainty (standard deviation) that the measured value matches the true value of the output distribution scales as $N^{-1/2}$. {Therefore the expected length of a computation itself influences the meaningful representational precision of the inputs; we found that our 4-bit programmable bitstream generator gave sufficient resolution at $N=128$.} The link between these two types of precision is subtle \cite{Manohar_2015} but will be increasingly important in determining when stochastic computing is or is not energy efficient for different target applications. 

\begin{figure}
\centering
\includegraphics[width=\columnwidth]{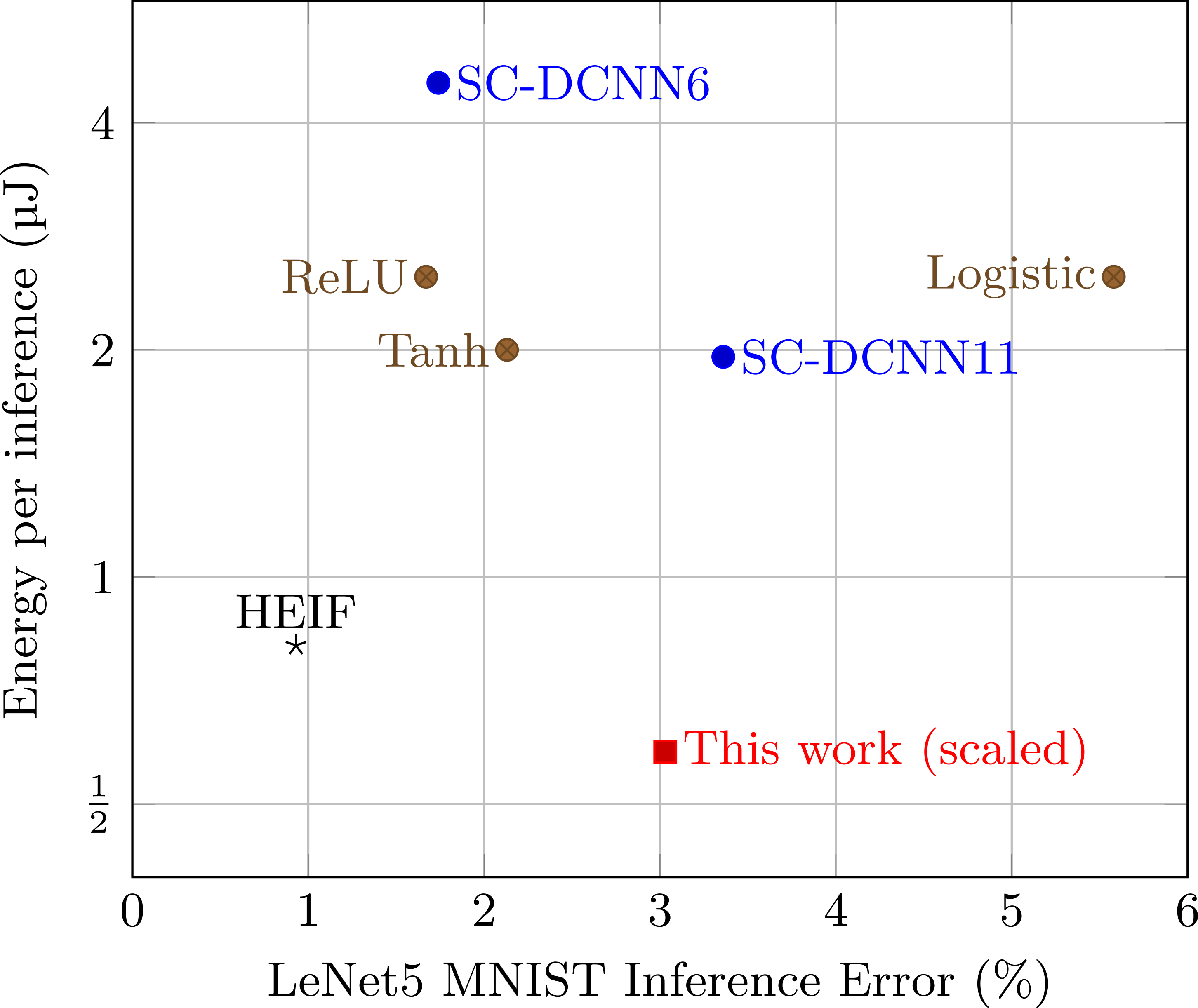}
\caption{Comparsion of stochastic computing implementations of LeNet5 {on the MNIST dataset} from the literature.  The survey of logistic, ReLU, and tanh networks is reported in Ref.~\cite{li2017hardware}. Two best-performing examples were extracted for SC-DCNN~\cite{Ren_2017}, and HEIF is reported in Ref.~\cite{Li_2018}. {These references are all reported at the 45~nm technology node.} The results for our work are presented at 1~V with $N=128$ and $\delta=16$ and are scaled up by a factor of four {to bring our 22~nm node calculation into fair comparison with the literature.}}
\label{fig:error-literature}
\end{figure}

{In Fig.~\ref{fig:error-literature}, we compare with the current state of the art in stochastic computing research. The works we cite all treat LeNet5 on MNIST. However, these works were also simulated at the 45~nm, rather than 25~nm, technology node. As is standard practice, we scale our reported energies by the corresponding scaling in transistor size (and therefore capacitance) to make a fair comparison~\cite{carballo2014itrs}. We found that with significant energy savings we could achieve comparable accuracy, 97~\%. While HEIF, the most modern work in that field, does significantly better (around 99.1~\%), it is also the culmination of a research program that optimized all the other networks in Fig.~\ref{fig:error-literature}, which originally performed at higher energy and lower performance. We have made no similar attempt to optimize our network and believe that it could be improved in principle to perform with similar accuracy and significantly lower energy.}

{The question remains as to whether 97~\% is a useful recognition accuracy. The answer is conditionally yes, as long as one chooses the correct application. Though our network addressed the “Hello, world!” task of handwriting recognition, one might imagine a similar convolutional neural network used for face recognition. Presumably, our implementation would not perform as well as what can be done in mainstream neural networks. For mission critical applications like biometric identification, this would be an unacceptable drop in performance. But for, say, automatic face detection in a power-constrained edge context like a mobile device’s camera application, the drop in performance may be well worth the considerable energy efficiency advantages.} 

\section{Conclusion}
\label{sec:conclusion}

We introduce a hybrid approach to classical stochastic computing, based on truly stochastic, low energy bitstreams generated by SMTJs. We introduce energy efficient primitive circuit elements (SC-PCSAs) that can be used to interface SMTJs with standard stochastic circuits. To test their effectiveness and explore the relaxed design space constraints afforded by true randomness, we simulated a neural network with \texttt{OR} gate neurons driven by SMTJ-based bitstream generators. Stochastic simulations of the network give 97~\% classification accuracy on the MNIST dataset and circuit simulations of the circuit elements show that the energy usage should be about 150~nJ per inference, {several times} less than other stochastic implementations of LeNet5. 

The energy efficiency we find in this case study is made possible in large part by true randomness, which eliminates the source correlations that would arise from the use of periodic pseudorandom number generators. Timing-based decorrelators called isolators address but do not entirely eliminate graph correlations, which arise due to reconvergent fanout of stochastic bitstreams. We analyze the energy and power usage of our neural network and discuss, in particular and in general, how SMTJ-based stochastic computing will scale with the progress being made in materials science. From the neural network perspective, our case study of LeNet5 was not optimized over all potential hyperparameters, so we expect that it should be possible to achieve higher classification accuracy and greater energy efficiency with more research. {Non-neuromorphic stochastic computing designs~\cite{alaghi2015strauss} may also be able to benefit from correlation-free randomness sources.} The design space degrees of freedom enabled by true randomness are an exciting landscape for future research.

\begin{acknowledgments}
  MWD and AMa contributed equally to this work. We thank Julie Grollier, Keven Garello, William Borders, Timothy Sherwood, George Tzimpragos, and Brian Hoskins for enlightening discussions and comments on the manuscript. MWD, AMa, PT, and AMi acknowledge support under the Cooperative Research Agreement Award No.~70NANB14H209, through the University of Maryland.
\end{acknowledgments}

\appendix

\section{The SR-latch in the SC-PSCA}
\begin{figure*}
\centering
\includegraphics[width=\textwidth]{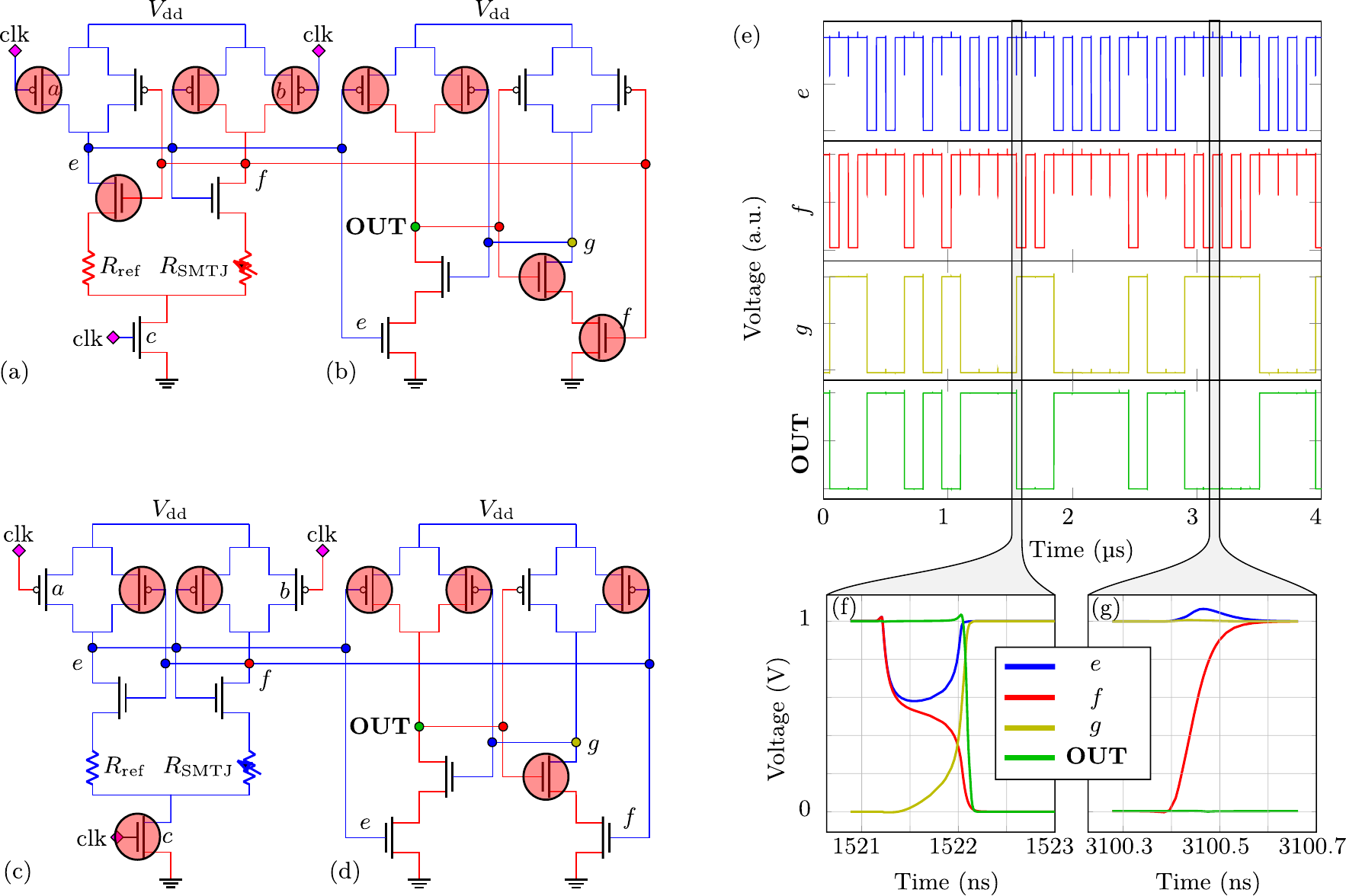}
    \caption{Operation of SC-PCSA.  Blue wires represent $V=V_\text{dd}$, red wires represent $V=0$, and red circles indicate that the source-drain path of transistor is turned off. (a): steady-state conclusion of evaluation phase, corresponding to the far-right of (f). The PCSA has detected that the SMTJ is in the parallel state, pulling node $f$ to ground and node $e$ to $V_\text{dd}$ as a result. (b): The SR-latch copies the value of node $f$ to the output node.
    (c): steady-state conclusion of reset/pre-charge, corresponding to the far-right of (g). (d): The PCSA has brought both $e$ and $f$ to $V_\text{dd}$ in preparation for the measurement in the next evaluation phase. Although $f$ has been brought to $V_\text{dd}$, the internal state of the latch does not change, so the output node continues to correctly represent the previously measured state of the SMTJ. (e, f, g): States of the various circuit node voltages throughout a simulation of the SC-PSCA. Curve colors correspond to points in the circuit indicated by the small circles in (a-d).}
    \label{fig:srlatch_dynamics}
\end{figure*}
\label{sec:srlatch}
Judiciously adding the set-reset (SR) latch is a crucial modification we make to the standard PSCA circuit. The SR latch is a standard electrical engineering concept that can be found in  textbooks~\cite{weste1994principles}. Because the operating principles of the SR latch may nevertheless be unfamiliar to some readers, we include in this appendix a brief description of the activity in the circuit in Fig.~\ref{fig:pcsa-circuit}. Typical descriptions of the SR-latch are given in terms of two recurrently connected \texttt{NAND} gates, but we will analyze it at the transistor level since the operation of a field-effect transistor is more physically transparent and familiar to many physicists.

We described in Sec.~\ref{sec:smtj-pcsa-readout} how nodes $e$ and $f$ are brought to either \texttt{01} or \texttt{10} after the clock signal goes to \texttt{1}, the so-called evaluation phase. It seems that the state of node $f$ is a good representative of the current state of the SMTJ, but unfortunately node $f$ goes to a high voltage state in the pre-charge phase when the clock goes back to \texttt{0}. Therefore $f$ carries artifacts of the circuit operation, and does not faithfully represent the physical state of the SMTJ at the time it was previously read. We fix this by adding the SR-latch in Fig.~\ref{fig:pcsa-circuit}(b). 

First, consider the conclusion of the evaluation phase for the case $e=\mathtt{1}$ and $f=\mathtt{0}$, as illustrated in Fig.~\ref{fig:srlatch_dynamics}(a,b). In this figure, blue wires are at $V_\text{dd}$ while red wires are connected to ground. The evaluated voltages at $e$ and $f$ are physically wired to the the $e$ and $f$ branches of the latch in Fig.~\ref{fig:srlatch_dynamics}(b). Because branch $e$ is on, the top-left transistor path is disconnected, denoted by a red circle. One can verify by hand that both transistors above the output node are disconnected (in the sense of open switches), while the transistors below the output node are connected to ground. The transistors above and below node $g$ are in the opposite configuration. The output node thus reliably captures the state of node $f$ during the evaluation phase. Notice that there are no paths for current to flow from $V_\text{dd}$ to ground, so ohmic losses terminate as soon as this steady state is reached.

Now we consider the pre-charge phase in Fig.~\ref{fig:srlatch_dynamics}(c,d). Both nodes $e$ and $f$ are brought to a high voltage state in preparation for the $RC$ race that will occur when the clock goes high again. Now that node $f$ is high, the top and bottom transistors change state on the $f$ branch of the latch, in the far right of Fig.~\ref{fig:srlatch_dynamics}(d). However, this does not change the state of the output node (or node $g$), node $g$ (and the output node) themselves contribute to the opening and closing of each other's paths to $V_\text{dd}$ and ground. Again, there is no steady state current flow once equilibrium is reached, so there are no continuous ohmic losses.

The two parallel transistors above node $g$ act like a \texttt{OR} gate; if either of their gate voltages is low, then a current path exists between $V_\text{dd}$ and node $g$. The series transistors below node $g$ act like an \texttt{AND} gate; a current path exists from $g$ to ground only if both gate voltages are high. This logical analysis will lead one to the standard presentation, but for our purposes it is sufficient to see that a change in state of $f$ does not cause a change of state in the output node, so the unwanted comb structure from Fig.~\ref{fig:SMTJ_wf}(d) is avoided.

\section{Comparison with stochastically-switched MTJs}
\label{app:switchedMTJs}
A popular method in the literature for random number generation based on magnetic nanotechnology is the use of stochastic switching phenomena in \emph{non-volatile} MTJs \cite{fukushima2014spin,choi2014magnetic,mondal2019energy}. In this approach, an MTJ is first written to one preferred configuration, and then a sub-critical write voltage is placed across an MTJ. The strength of the write voltage is related to the parameter $p$ of the Bernoulli trial to be performed. After probabilistically writing to the MTJ, the state is sensed with a read voltage in the usual way; the MTJ will have switched to the other stable configuration with probability $p$.

To compare our method with this one, we refer to Ref.~\cite{choi2014magnetic} as a standard point of reference. Though that reference does not report numbers for the entire circuitry, they do report resistance and voltage numbers for the tunnel junction itself. Their devices have an average resistance $R= 1000~\Omega$, and they apply an average of $217.5~\text{mV}$ bias perturbative write voltages over an average window of $8.75~\text{ns}$. We neglect the cost of their read operation and any required CMOS circuitry here; the ohmic losses in the MTJ alone amount to 414~fJ per bit. Ref.~\cite{mondal2019energy} uses similar techniques on MTJs specifically tailored to the generation of stochastic computing bitstreams (the same task we consider here); they report average costs of 526~fJ per bit.

Though the energy per bit is orders of magnitude higher than what we have presented in Fig.~\ref{fig:energy-nums}, the tasks are not immediately equivalent. The main difference seems to be in speed available in the current state-of-the-art. Write/read cycles on non-volatile MTJs can be quite fast, allowing for rapid generation of random numbers in principle. Our scheme and that of \cite{PhysRevApplied.8.054045} avoid large ohmic losses by using sense amplifiers, but as a result are limited in speed by the natural time scale of the SMTJ's random telegraph noise. Experiments on SMTJs have demonstrated millisecond scale fluctuations; uncorrelated random bits cannot be sampled significantly faster than this time scale, which goes as roughly $e^{\Delta/k_BT}~\text{ns}$. If faster SMTJs can be engineered, our scheme would continue to function at the same energy per bit performance that we describe in this paper, and could plausibly produce bits down at the nanosecond time scale for ultra-low barrier devices.

\section{Comparison with $p$-bits}
\label{app:p-bits}
A common application of SMTJs is the $p$-bit \cite{faria2018implementing,jia2018spintronics,sengupta2016probabilistic,PhysRevX.7.031014,ganguly2017reservoir,hassan2019voltage, shim2017ising}. A $p$-bit consists of an SMTJ and a transistor in series between a supply voltage $V_\text{dd}$ and ground. The node between the SMTJ and the transistor is a voltage divider; its output is sent to a sequence of specially tuned CMOS inverters, where that output fluctuates around some reference voltage to produce a random telegraph noise signal. Controlling the gate-source voltage of the transistor in an analog fashion allows one to access analog tunability of the spin-torque on the SMTJ; this consequently allows one to tune the probability encoded in the output of the CMOS inverters.

Generally speaking, ideal $p$-bits are in a different class of device than what we propose here. In particular, they operate on analog voltage inputs. One could in principle build an analog probabilistic computer out of these units, whereas our proposed circuit is purely digital. Here, we do not compare the relative computational ability of digital versus analog computing but evaluate how well $p$-bit circuits would work for stochastic computing.

To make a fair comparison of device power and energy, we assume the supply voltage is again $V_\text{dd}=$~1~V. The voltage division provided by the inverter and SMTJ in the $p$-bit swings between two values which must be distinguished by the CMOS inverter. In order to center the $p=1/2$ response of the $p$-bit at an input voltage of $V=0.5~\text{V}$, the effective resistance of the transistor at that input voltage should be $\sqrt{R_P R_{AP}}$ to maximize the sensitivity of the circuit to both the high and low resistance states. In that configuration, we essentially have a $1~\text{V}$ voltage drop across a series resistor of effective resistance $\sqrt{R_P R_{AP}}+(R_P+R_{AP})/2$. The current running through the structure is therefore
\begin{equation}
    I = \frac{2V_\text{dd}}{(R_P+R_{AP})+2\sqrt{R_P R_{AP}}}.
\end{equation}
For $R_{AP} = 100~\text{k}\Omega$ and $R_P = 50~\text{k}\Omega$, this amounts to a 3.4~{\textmu}A current, or 3.4~{\textmu}W of ohmic dissipation per device. Neglecting the energy cost associated with generating independently controllable analog voltage sources for each $p$-bit, the ohmic loss is about 500~fJ per bit at the 150~ns clock cycle we consider, substantially greater than the $\approx$~10~fJ for our PCSA-based approach.  However, as the clock speed of the circuit is increased, the PCSA-based approach has constant energy cost per bit while the $p$-bit approach scales roughly linearly to lower energies with increasing speed. In order to achieve 10~fJ per bit performance, the autocorrelation time of the stochastic fluctuations in a $p$-bit would need to be reduced to 3~ns. 

Recent work has attempted to push the theory of SMTJs toward this limit where the analog behavior can be harnessed more energy efficiently. Ref.~\cite{borders2019integer} makes similar order of magnitude energy projections as ours in the case that the SMTJ dwell time could be reduced to $\approx 1~\text{ns}$. Theory~\cite{kaiser2019ultrafast} suggests that nanomagnets with autocorrelation times on this scale might be realizable in the limit that the barrier goes to zero. Realizing devices that operate in this regime face a number of obstacles. As the barrier goes to zero, the current becomes more and more efficient at dictating the SMTJ state, requiring a delicate balance between adequate read currents and currents that control the state of the nanomagnet. Fabricating such a device may require extremely narrow margins of error.

\section{Device variability}
\label{app:variability}

\begin{figure}
  \centering
  \includegraphics[width=\columnwidth]{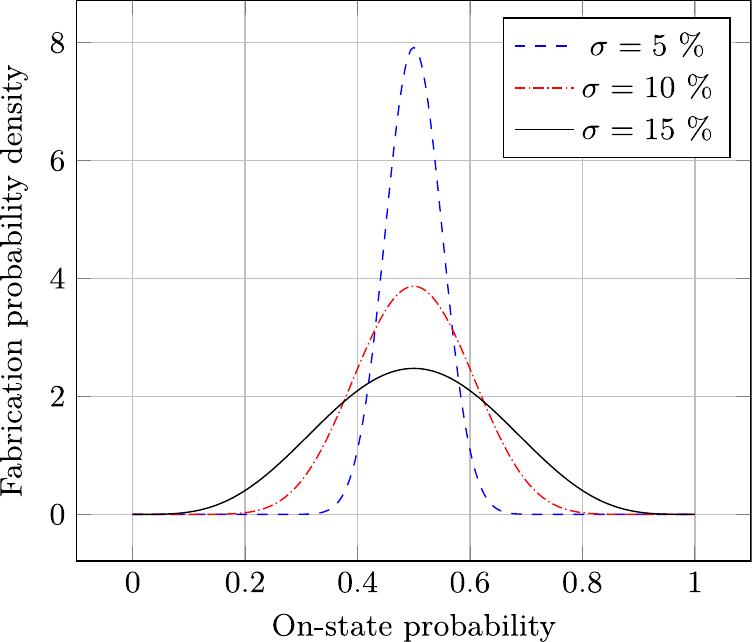}
  \caption{Probability density functions for the beta distribution centered at $1/2$ with three different standard deviations.}
  \label{fig:beta-dists}
\end{figure}

In the main text, especially leading up to Fig.~\ref{fig:distribution-plots}, we assume a distribution for device variability. However, we do not assume the distribution to be gaussian, because the exponential tails of the normal distribution are problematic; the probability that a particular device is found in the parallel or antiparallel states is \emph{strictly} confined to the open subset $(0,1)$, by construction. Instead of trying to truncate or morph normal distributions to the domain, we opt instead to use the beta distribution, which is naturally defined on the unit interval.

Suppose that the probability for a device to be in the on-state is itself
a random variable $\mathcal P$. We choose to model $\mathcal P$ as a beta
distribution with unnormalized probability density
\begin{equation}
    f(p)\propto \begin{cases}
    p^{\mu\phi-1}(1-p)^{\phi-\mu\phi-1} & 0 \leq p \leq 1 \\
    0 & \text{otherwise}.\end{cases}
\end{equation}
The normalization is given by an Euler integral of the first kind. A convenient parametrization of the beta distribution is to specify its mean $\mu$ and a shape parameter $\phi>0$, which is related to the variance of the distribution by
\begin{equation}
  \label{eq:variance}
  \sigma^2 = \frac{\mu(1-\mu)}{1+\phi}.
\end{equation}
First, consider the case where $\mu=1/2$ but $\phi$ is greater than zero. We plot three different distributions with standard deviations $\sigma\in\{0.05,0.1,0.15\}$ in Fig.~\ref{fig:beta-dists}. The existence of these distributions at the inputs to the bitstream generator will induce distributions in the statistics of the generator's output bitstream. Expressions for these output distributions are analytically tractable~\cite{springer1970distribution}, but are given in terms of hypergeometric Meijer G-functions and provide little useful intuition. To gain insight into the relationship between device variance and bitstream variance, we sample five thousand generator outputs at each programming code, for each of the distributions in Fig.~\ref{fig:beta-dists}. Fig.~\ref{fig:distribution-plots} plots these distributions vertically at each programmable probability. 

A similar idea for producing a discretely-specified probability bitstream from a collection of LFSRs, called the weighted binary generator (WBG), has been discussed in Refs.~\cite{gupta1988binary,yang2018towards}. Our circuit has the advantage of $O(n)$ scaling of the input capacitance as a function of bit resolution $n$, whereas the input capacitance of the WBG scales as $O(n^2)$. We expect our solution to take less energy in general, regardless of whether LFSRs or SMTJs are used as the randomness sources. On the other hand, our numerical experiments indicate that for large deviations of the mean input probabilities from $1/2$, distortions in the output behavior of the WBG are more well-behaved than ours in the sense that the ordering of outputs is preserved relative to the WBG's binary programming. In the case that programming protocols can be determined after fabricated devices are characterized, our circuit remains favorable. But if the mean of beta distribution governing expected probability for each SMTJ is unknown, or if the SMTJ variance is very large and programming proctocol needs to be uniform across many devices, then the use of the WBG circuit may be preferable for reliability and ease of programming.

\section{Details of the neural network}
\label{app:nn-details}
\subsection{Layer structures}
\begin{figure}
    \centering
    \includegraphics[width=\columnwidth]{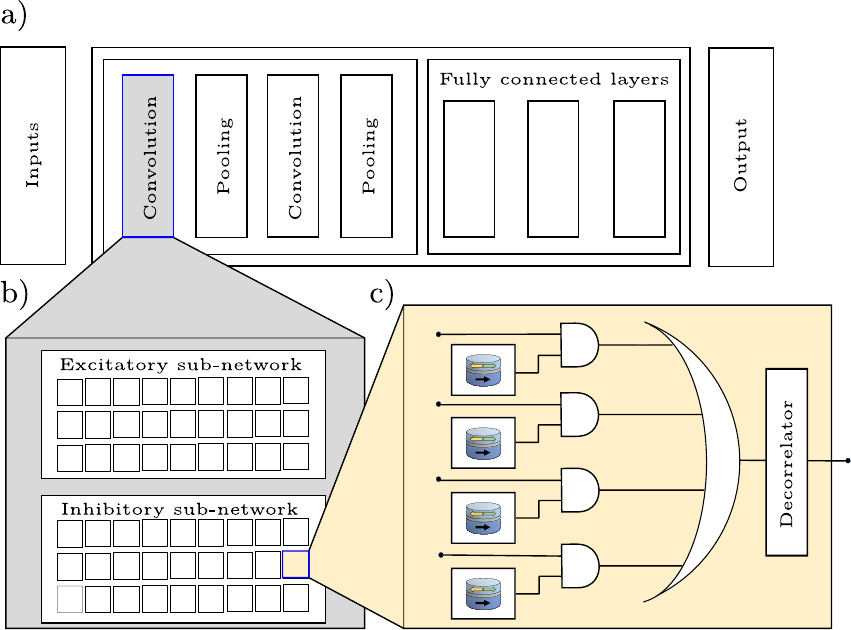}
    \caption{(a) Schematic block diagram of LeNet5 neural network. (b) Schematic block diagram of a neural layer (either a convolutional or a fully connected layer). Each of these layers are composed by an excitatory and inhibitory sub-networks. (c) Schematic view of a functional neural block comprised of superparamagnetic tunnel junctions, \texttt{AND} gates, one \texttt{OR} gate, and a decorrelator.}
\label{fig:layout}
\end{figure}

Each layer in a stochastic neural network architecture receives output bitstreams from the previous layer, as well as bitstreams from the programmable SMTJ weight arrays. These bitstreams are all multiplied in parallel using  \texttt{AND} gates, the outputs of which are fed into the \texttt{OR} gate neurons for summation and activation. In other words, a fully connected layer with input degree $m$ and output degree $n$ is simply implemented as $n$ different $m$-input \texttt{OR} gate neurons as we described in Sec.~\ref{sec:neuralnetworks}. The output of each layer is then passed through a random mask decorrelator as we described in Appendix~\ref{sec:decorr-arch}.

In a convolutional layer, we spatially multiplex the kernel applications. This means that each weight, which is being generated as described in Sec.~\ref{sec:smtj-pcsa-readout}, is fanned out to a large number of input pixels at the same time. This massive weight sharing will cause neighboring pixels in the layer output to be strongly correlated, making the use of a graph decorrelator especially important here. 

Today, most neural network designers implement the $\max$ operation as their pooling function, though average-pooling and min-pooling are also used. Ref.~\cite{neugebauer-2019} has developed a correlation-insensitive max function for stochastic circuits that could be used for this purpose. 
Our pooling layer performs ``\texttt{OR}-pooling''---that is, we simply use a single 4-input \texttt{OR} gate as our $2\times2$ pooling operation. This is similar to the method used in Ref.~\cite{li2017neural} to approximate average pooling, but without their weighting elements.

\subsection{Dual architecture}
\label{sec:dual-architecture}

In this paper, we use unipolar encoding to represent values in the stochastic circuit. To use unipolar encoding as such is to identify the probability of a wire being in the on state with the value that wire is said to encode. A disadvantage of this approach is that only numbers between zero and one can be encoded by the network. An alternative is bipolar encoding, where a wire turned on with probability $p$ is said to encode the value $2p-1$~\cite{winstead2019tutorial}. This approach is used by many others in the field~\cite{Li_2018,Ren_2017}, but unfortunately it is not clear to us that the \texttt{OR} gate can still be used as a useful nonlinear activation function in this encoding scheme. 

Inhibitory behavior is generally believed to be a crucial aspect of neural networks~\cite{chialvo1999learning,bak2001adaptive}. In most previous stochastic network proposals, use of the bipolar representation \cite{winstead2019tutorial} grants the network access to negative numbers and, consequently, a mechanism for inhibitory signals. Some work has also been done on learning in networks with nonnegative weights~\cite{chorowski2015learning}, where inhibition is exercised by amplifying all signals except for those targeted for inhibition. Unfortunately, the unipolar representation employed here can represent neither negative numbers nor numbers greater than unity. Unipolar stochastic networks therefore lack the usual inhibitory mechanisms and fail to meaningfully learn most nontrivial data sets in our experiments. 

To address this issue, we use a variation on a method proposed in Refs.~\cite{Tomlinson_1990,Young_Chul_Kim_1995_architecture}. We refer to our strategy as a \emph{dual architecture}. We employ two separate weight matrices in each layer, which are labeled as the excitatory and inhibitory subnetworks. The entire output vector from the inhibitory subnetwork is then elementwise inverted and AND'd with the excitatory output to achieve the response of the full layer. In the absence of correlations, the $j^\text{th}$ element $z_j$ of the final output from a network layer is therefore given by
\begin{equation}
\label{eq:subnetwork-eq}
    z_j = y^{(e)}_j \left(1-y^{(i)}_j\right),
\end{equation}
where $y_j^{(e,i)}$ is the output element for the excitatory (inhibitory) subnetwork. {This leads more explicitly to Eq.~\eqref{eq:or-and-nor} from the main text.} In reality, the preliminary outputs $y_j^{(e)}$ and $y_j^{(i)}$ are never uncorrelated, because they are both sourced from the same set of inputs $\{x_j\}$ to the layer. However, the use of the approximation in Eq.~\eqref{eq:subnetwork-eq} is necessary to use standard, local backpropagation methods, and we find that any induced error is minimal.

\subsection{Simulation and Training}

We build two kinds of models for our architecture: an analytic one and a stochastic one. The analytic model provides a representation of the network in terms of probabilities rather than the more complicated stochastic representation of random processes, allowing it to run faster than the stochastic model. The analytic model verifies the functional correctness of our network structure and provides a baseline against which to compare our stochastic simulations. The stochastic model is a logic-level simulator, which allows us to determine the effects of correlation and to estimate activity factors needed for energy estimates. It uses input bitstreams that are generated from statistics consistent with the known properties of SMTJs.

To train our network, we use the standard backpropagation algorithm applied to our nonstandard network functionality given by {Eq.~\eqref{eq:or-and-nor} (that is, Eqs.~\eqref{eq:or-gate} and~\eqref{eq:subnetwork-eq}).}
The speed of the analytic model makes it suitable for the training phase, as running detailed stochastic simulations for each inference of the training process would be prohibitive. This model is deterministic in the sense that we use the analytic probability equations for inference and backpropagation; it is local in the sense that we assume the inputs to each node of the network to be free of correlations. 
To train the model, we use a mini-batch version of the RMSProp optimization algorithm~\cite{hinton2012neural}. We use a learning rate of 0.005 and a forgetting factor of 0.95. We train sixty randomly initialized models for sixteen epochs each and select the best performing model. This network is then transferred to the stochastic model, to be simulated using realistic bitstreams generated according to known statistics for physical SMTJ devices.

\subsection{Decorrelation}
\label{sec:decorr-arch}
In order to train our neural network efficiently, we use backpropagation and gradient descent. In practice, backpropagation is implemented as a local learning rule; the gradient of the cost function is determined at each node in the network as a function only of that node and its nearest, connected neighbors. This locality is essential in keeping backpropagation algorithmically efficient, and therefore necessitates the use of a local analytic description that assumes all inputs are statistically independent. Such an analytic description, however, will necessarily fail to capture most graph correlations. In the specific case of a neural network, then, we need additional functionality for addressing graph correlations.

Because of the massive fan-in required in neural network systems, the algorithm from Ref.~\cite{Ting_2016} would be unwieldy to implement in a neural network. For a fully connected layer with $n$ neurons on the output, total decorrelation would require different delays for each neuron, leading to $O(n^2)$ delay elements per neuron. Our approach is to instead delay each neuron by a pseudorandom but fixed amount, creating a random mask of delay lengths on the output of each neural network layer. Such a programmable feature is implemented in our architecture by a chain of flip flops which are tapped into a multiplexer. The output of each decorrelator is fed to the next layer. In other words, every output of every layer is delayed by some integer between zero and a fixed upper bound, chosen pseudorandomly and uniformly over the interval at the network programming step.  This approach does not formally eliminate graph correlations, but it empirically reduces their impact to a low enough level such that implementation of the network architecture becomes feasible. The results presented in Sec.~\ref{sec:eval} demonstrate that the delay time can be set long enough to mostly saturate correlation-based network errors.

\bibliography{references}

\end{document}